# Generalized Chaotic Synchronization in Coupled Ginzburg–Landau Equations

A. A. Koronovskiĭ, P. V. Popov, and A. E. Hramov

*Saratov State University, Saratov, 410012 Russia*
*e-mail: alkor@nonlin.sgu.ru; popovpv@nonlin.sgu.ru; aeh@nonlin.sgu.ru*

**Abstract**—Generalized synchronization is analyzed in unidirectionally coupled oscillatory systems exhibiting spatiotemporal chaotic behavior described by Ginzburg–Landau equations. Several types of coupling between the systems are analyzed. The largest spatial Lyapunov exponent is proposed as a new characteristic of the state of a distributed system, and its calculation is described for a distributed oscillatory system. Partial generalized synchronization is introduced as a new type of chaotic synchronization in spatially nonuniform distributed systems. The physical mechanisms responsible for the onset of generalized chaotic synchronization in spatially distributed oscillatory systems are elucidated. It is shown that the onset of generalized chaotic synchronization is described by a modified Ginzburg–Landau equation with additional dissipation irrespective of the type of coupling. The effect of noise on the onset of a generalized synchronization regime in coupled distributed systems is analyzed.

## 1. INTRODUCTION

Synchronization of chaotic oscillations is one of the most important fundamental nonlinear phenomena [1–4]. On the one hand, extensive studies in this area improve theoretical understanding of the general laws governing the interactions between complex nonlinear systems of different (chemical, physical, biological etc.) nature [1, 5–8]. On the other hand, the practical applications of these studies include information transmission technologies using deterministic chaotic oscillations [9, 10], dynamics of neural ensembles [11, 12], disease diagnostics [13, 14], etc.

Studies of chaotic synchronization have mainly relied on analysis of coupled low-dimensional dynamical systems (lumped-parameter models of flow systems or maps) [1–3, 5]. Currently, different types of synchronous behavior of chaotic systems are distinguished: phase synchronization [2, 16, 17], generalized synchronization [18, 19], lag synchronization [20], intermittent lag [21] and generalized [22] synchronization, noise-induced synchronization [23–25], complete synchronization [26–29], and time-scale synchronization [30–32]. All of these are related to each other (e.g., see [25, 31–35]). However, interrelations between them are not fully understood. Each type of synchronous behavior is detected by specific methods for analyzing coupled oscillatory systems (e.g., see [1, 2]).

Current research is increasingly focused on complex dynamics of distributed oscillatory systems exhibiting spatiotemporal chaotic behavior, pattern formation, etc. [36–42]. At the same time, extensive studies are being conducted on the feasibility of control and synchronization of spatiotemporal chaotic oscillations in chains, lattices, and networks of coupled oscillators [1, 43–45], as well as in distributed oscillatory systems. In particular, ongoing research deals with control and synchronization in distributed oscillatory systems described by Ginzburg–Landau [1, 46–50] and Kuramoto–Sivashinsky [51] equations, oscillatory chemical reactions [5, 52], and beam–plasma and electron–wave systems in plasmas [53, 54].

Most analyses of spatially distributed continuous oscillatory systems described by partial differential equations have been focused on complete chaotic synchronization of identical and slightly different coupled systems. Complete chaotic synchronization means that the state vectors of the coupled chaotic dynamical systems are equal, $\mathbf{x}_1(t) \approx \mathbf{x}_2(t)$ [26–29]. This regime arises in synchronized identical chaotic oscillators. When the corresponding control parameters of the systems are slightly different, the state vectors of the chaotic oscillators are nearly equal, $|\mathbf{x}_1(t) - \mathbf{x}_2(t)| \approx 0$, but remain different. Let us briefly discuss the main results of studies of chaotic synchronization of spatially distributed continuous oscillatory systems.

In [50, 51, 55, 56], the Ginzburg–Landau and Kuramoto–Sivashinsky equations were used as examples to demonstrate that complete synchronization can be achieved when various types of coupling is introduced between distributed chaotic systems, including spatially uniform and "pinning" couplings. In the latter case, spatially distributed systems are coupled only at certain points in space (typically, these points make up a periodic array with spatial period $\Delta L$) [43, 57]. The possibility of message encoding and efficient multichannel communication by means of complete chaotic synchronization in a distributed oscillatory medium was discussed in [58].

Phase synchronization of coupled complex Ginzburg–Landau equations was analyzed in [59]. Phase synchronization was defined by the condition $|\phi_{u_1}(x, t) - \phi_{u_2}(x, t)| < \text{const}$, where $\phi_u(x, t) = \arg u(x, t)$ denotes the phases of the coupled complex fields $u_1(x, t)$ and $u_2(x, t)$ described by the Ginzburg–Landau equations. In [54, 60], we showed that time-scale synchronization can be achieved in unidirectionally coupled distributed electron–wave systems with backward waves exhibiting spatiotemporal chaotic behavior [61].

Another highly important and extensively studied type of chaotic synchronization is generalized synchronization of unidirectionally coupled chaotic systems [18]. In generalized synchronization regimes, the state vectors of the driving and response systems, $\mathbf{x}_d(t)$ and $\mathbf{x}_r(t)$, are related by a functional relation $\mathbf{x}_r(t) = \mathbf{F}[\mathbf{x}_d(t)]$ after the initial transients have died out (which may take a long time [22]). This relation may be very complicated (e.g., it may have a fractal structure [19]), and its explicit form cannot be found in most cases. Several methods for detecting generalized synchronization between unidirectionally coupled chaotic systems [19, 62].

It should be noted that generalized chaotic synchronization was studied in detail for low-dimensional flow systems and maps [18, 19, 62]. In [63, 64], we analyzed the physical mechanisms underlying the onset of generalized chaotic synchronization and showed that the behavior of the response chaotic system in a generalized synchronization regime is equivalent to the behavior of a modified system with additional dissipation under the external forcing by the chaotic signal generated by the driving system.

However, generalized synchronization of spatially distributed oscillatory systems exhibiting spatiotemporal chaotic behavior has never been studied in detail. We can only refer to [5], where it was demonstrated that generalized synchronization was achieved in a chemical reaction model. Therefore, analysis of the onset of generalized chaotic synchronization regimes in distributed oscillatory and continuous systems is of great fundamental interest.

In this paper, we analyze generalized chaotic synchronization in unidirectionally coupled oscillatory media described by Ginzburg–Landau equations with different types of unidirectional coupling between the driving and response systems. Our choice of the spatially distributed systems described by Ginzburg–Landau equations is motivated by the fact that these systems are used as reference models in studies of spatiotemporal chaos and pattern formation in various distributed media [2, 43, 46].

## 2. MODELING AND DETECTION OF GENERALIZED SYNCHRONIZATION

The mathematical model of distributed oscillatory media analyzed in this study is the system of two unidirectionally coupled one-dimensional Ginzburg–Landau equations for complex fields $u(x, t)$ and $v(x, t)$ [46],

$$\frac{\partial u}{\partial t} = u - (1 - i\alpha_d)|u|^2 u + (1 + i\beta_d)\frac{\partial^2 u}{\partial x^2}, \quad (1)$$
$$x \in [0, L],$$

$$\frac{\partial v}{\partial t} = v - (1 - i\alpha_r)|v|^2 v + (1 + i\beta_r)\frac{\partial^2 v}{\partial x^2} + \varepsilon H(t - \tau)\mathcal{F}[u, v], \quad (2)$$
$$x \in [0, L],$$

subject to the periodic boundary conditions

$$u(x, t) = u(x + L, t), \quad v(x, t) = v(x + L, t), \quad (3)$$

where $L$ is the spatial period of the system.

Equations (1) and (2) describe driving (autonomous) and response distributed systems, respectively. (Recall that the concept of generalized synchronization is applied only to systems with unidirectional coupling.) In the coupling term, $\mathcal{F}[u, v]$ is called *coupling function*, the coupling strength is quantified by $\varepsilon$, and $H(\eta)$ is the Heaviside step function ($H(\eta) = 0$ at $\eta < 0$ and $H(\eta) = 1$ otherwise; i.e., the distributed systems are uncoupled during the transient time $\tau$).

Unless specified otherwise, the values of the control parameters $(\alpha_d, \beta_d)$ and $(\alpha_r, \beta_r)$ of the driving autonomous) and response systems are set as follows: $\alpha_d = 1.5$, $\beta_d = 1.5$, $\alpha_r = 4.0$, and $\beta_r = 4.0$. The spatial period of the system is $L = 40\pi$. These values of control parameters correspond to spatiotemporal chaotic behavior of autonomous distributed systems [46, 56].

We computed Eqs. (1) and (2), starting from random initial conditions $u(x, t = 0)$ and $v(x, t = 0)$ for the complex fields and using an explicit second-order accurate finite-difference scheme with $\Delta t = 0.0002$ and $\Delta x = L/1024$.

To detect generalized synchronization in distributed systems, we apply the auxiliary system method originally proposed in [62] for low-dimensional systems. In this method, an auxiliary system with state vector $\mathbf{x}_a(t)$ identical to the response system with state vector $\mathbf{x}_r(t)$ is considered. The initial state $\mathbf{x}_a(t_0)$ of the auxiliary system is different from the initial state $\mathbf{x}_r(t_0)$ of the response system, but lies in the same basin of attraction. When the coupled systems are not synchronized, $\mathbf{x}_r(t)$ and $\mathbf{x}_a(t)$ are different state vectors belonging to the same chaotic attractor. In a generalized synchronization regime, since the relation $\mathbf{x}_r(t) = \mathbf{F}[\mathbf{x}_d(t)]$ entails $\mathbf{x}_a(t) =$

$\mathcal{F}[\mathbf{x}_d(t)]$, the states of the response and auxiliary systems must become identical ($\mathbf{x}_r(t) \equiv \mathbf{x}_a(t)$) after the initial transients have died out (which may take a long time [22]). Thus, the equivalence of the states of the response and auxiliary systems after an initial transient time provides a criterion for detecting generalized synchronization of the driving and response chaotic systems.

The auxiliary system must also be described by a nonautonomous Ginzburg–Landau equation (cf. Eq. (2)):

$$\frac{\partial v_a}{\partial t} = v_a - (1 - i\alpha_r)|v_a|^2 v_a + (1 + i\beta_r)\frac{\partial^2 v_a}{\partial x^2} + \varepsilon H(t - \tau)\mathcal{F}[u, u_a], \quad (4)$$
$$x \in [0, L].$$

This equation is supplemented with initial conditions that are different from those for the response oscillatory system:

$$v_a(x, t = 0) \neq v(x, t = 0). \quad (5)$$

Under these conditions, if $v_a(x, t) = v(x, t)$ after the coupling is switched on at $t = \tau$, then one can conclude that unidirectionally coupled distributed systems are in a generalized synchronization regime.

Alternatively, generalized synchronization can be detected by calculating the largest conditional Lyapunov exponent for unidirectionally coupled systems [65]. This method can also be applied to detect generalized chaotic synchronization of low-dimensional systems [19]. In this method, Lyapunov exponents are calculated for the response system. Since its behavior depends on the dynamics of the driving system, these Lyapunov exponents are different from those for the autonomous response system and are referred to as *conditional*. A criterion for generalized synchronization in unidirectionally coupled dynamical systems is formulated as the requirement of a negative largest conditional Lyapunov exponent: $\lambda_{c1} < 0$ [19].

To apply a similar method to a spatially distributed system, one has to calculate a characteristic analogous to the Lyapunov exponent for low-dimensional dynamical systems. In the next section, we propose *spatial Lyapunov exponent* as a new measure of chaoticity of spatiotemporal dynamics of a distributed system and develop a method for calculating this exponent, based on a modification of Bennetin's algorithm for low-dimensional systems [66, 67].

## 3. CALCULATION OF THE LARGEST SPATIAL LYAPUNOV EXPONENT

The largest Lyapunov exponent is a very important characteristic that quantifies the degree of chaoticity of a dynamical regime of a low-dimensional dynamical system [3]. Characterization of behavior of spatially distributed systems in terms of an averaged rate of exponential divergence of nearby trajectories (i.e., an analog of the Lyapunov exponent for low-dimensional systems) is a very attractive and promising idea. Attempts at adapting and extending the concept of largest Lyapunov exponent to analysis of dynamics of distributed systems usually amount to calculations of Lyapunov exponents by methods developed for lumped-parameter models. In particular, the largest Lyapunov exponent can be evaluated by processing a time series obtained from the signal measured at a point in the state space of a distributed system (e.g., see [67]), as done for lumped-parameter models [68, 69]. Alternatively, the Lyapunov spectrum can be calculated by applying Bennetin's algorithm [66] to a finite-difference counterpart of the distributed system. In the latter method, the discretized distributed system is treated as a high-dimensional dynamical system, and only a few largest exponents are evaluated in its Lyapunov spectrum (e.g., see [70]).

In this paper, the degree of chaoticity of spatiotemporal dynamics of the distributed system under analysis is quantified by introducing *spatial Lyapunov exponent*.

Denoting the state of the spatially distributed system at an instant $t_0$ by $\mathbf{R}(\mathbf{x}, t_0)$, we define the distance $s(\mathbf{R}_1, \mathbf{R}_2)$ between states as

$$s(\mathbf{R}_1, \mathbf{R}_2) = \sqrt{\int_V \|\mathbf{R}_1(\mathbf{x}) - \mathbf{R}_2(\mathbf{x})\|^2 dV}, \quad (6)$$

where $V$ is the volume of the system. We consider the evolution of the system from an "unperturbed" state $\mathbf{R}^0(\mathbf{x}, t_0)$ and from a perturbed state $\tilde{\mathbf{R}}^0(\mathbf{x}, t_0) = \mathbf{R}^0(\mathbf{x}, t_0) + \tilde{\xi}(\mathbf{x})$, where $\tilde{\xi}(\mathbf{x})$ is a random function, assuming that $s(\mathbf{R}^0, \tilde{\mathbf{R}}^0) = \epsilon$ is small.

Suppose that the vectors $\mathbf{R}_1(\mathbf{x})$ and $\tilde{\mathbf{R}}_1(\mathbf{x})$ representing the unperturbed and perturbed states at the instant $t_0 + T$ have been found by solving the governing equations. We renormalize the perturbed state $\tilde{\mathbf{R}}_1(\mathbf{x})$ so that its deviation from $\mathbf{R}_1(\mathbf{x})$ equals its starting value $\epsilon$:

$$\tilde{\mathbf{R}}_1^0(\mathbf{x}) = \epsilon\tilde{\mathbf{R}}_1/S(\mathbf{R}_1, \tilde{\mathbf{R}}_1).$$

After the renormalization is repeated $M$ times, the initial perturbation amplitude will be multiplied by the factor

$$P = \prod_{k=1}^{M} \frac{S(\mathbf{R}_k, \tilde{\mathbf{R}}_k)}{\epsilon}.$$

The largest spatial Lyapunov exponent $\Lambda$ is then defined as

$$\Lambda = \frac{1}{MT}\ln P = \frac{1}{MT}\sum_{k=1}^{M} \ln\frac{S(\mathbf{R}_k, \tilde{\mathbf{R}}_k)}{\epsilon}. \quad (7)$$

It is shown below that the degree of chaoticity of a spatially uniform system is adequately evaluated by using this quantity, whereas the proposed algorithm must be modified in the case of a spatially nonuniform system (see discussion below).

The state of a system described by the Ginzburg–Landau equation is represented by the vector $\mathbf{R}(\mathbf{x}) = \{\operatorname{Re} u(x), \operatorname{Im} u(x)\}$, and the corresponding integral in (6) is calculated over $x \in [0, L]$ (spatial period of the system). When unidirectionally coupled systems are governed by Eqs. (1) and (2), the largest spatial Lyapunov exponent $\Lambda_c$ is calculated for system (2) by treating the signal generated by the driving system as an external force.

Figure 1 illustrates the dependence of $\Lambda$ on the control parameters calculated for autonomous Eq. (1). One can see that $\Lambda$ increases with $\beta_d$. It can readily be shown that the solution to Ginzburg–Landau equation (1) is unstable if

$$\alpha_d \beta_d > 1. \tag{8}$$

Numerical calculations of the largest Lyapunov exponent are in good agreement with analytical result (8). In particular, $\Lambda$ becomes positive (transition to chaos occurs) as $\beta_d$ exceeds $1/\alpha_d$ for each $\alpha_d$, and the value of $\Lambda$ increases further with $\beta_d$. When $\beta_d < 1/\alpha_d$, we obtain $\Lambda = 0$, which corresponds to periodic oscillation.

## 4. SPATIALLY UNIFORM COUPLING

First, we consider the case of spatially uniform unidirectional diffusion coupling between distributed oscillatory systems:

$$\mathcal{F}[u, v] = u - v. \tag{9}$$

We examine the behavior of the system described by Eqs. (1) and (2) with coupling function (9) for several values of the coupling strength $\varepsilon$, while the values of control parameters $\alpha_d$, $\beta_d$, $\alpha_r$, and $\beta_r$ specified in Section 2 are held constant.

The top and bottom panels in Fig. 2 illustrate, respectively, the spatiotemporal behavior of the amplitude difference between the driving and response systems, $|v(x, t) - u(x, t)|$, and between the response and auxiliary systems, $|v(x, t) - v_a(x, t)|$, calculated for values of $\varepsilon$ increasing from left to right.

When $\varepsilon$ is relatively small, no general chaotic synchronization is observed (Fig. 2a, $\varepsilon = 0.5$): the coupled distributed systems oscillate differently during the computational time interval, and so do the $v(x, t)$ and $v_a(x, t)$ oscillations. The onset of a general chaotic synchronization regime is observed as $\varepsilon$ is increased to $\varepsilon_{GS} \approx 0.73$. In this regime, the driving and response systems described by (1) and (2) exhibit different spatiotemporal oscillatory behavior ($|u - v|$ does not vanish), whereas $|v - v_a| = 0$ after a transient time has elapsed (Fig. 2b, $\varepsilon = 0.9$), which implies a functional relationship between the states of the driving and response systems and, therefore, signifies the onset of a general synchronization in distributed systems.

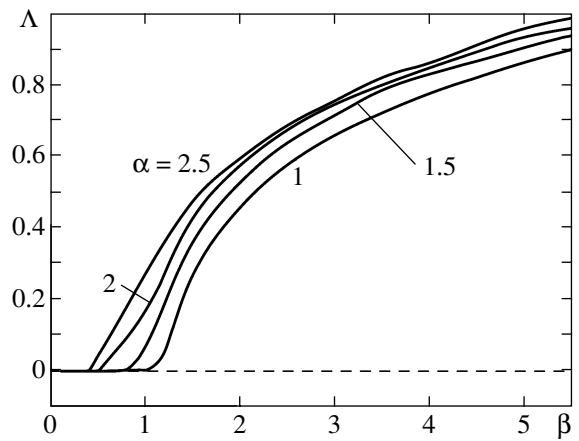

**Fig. 1.** Largest spatial Lyapunov exponent for autonomous Ginzburg–Landau (1) vs. parameter $\beta$ for several values of $\alpha$.

Complete chaotic synchronization of the coupled systems is attained with further increase in $\varepsilon$: zero difference is observed both between the response and auxiliary systems and between the response and driving ones (Fig. 2c, $\varepsilon = 20$). Thus, complete chaotic synchronization in coupled distributed systems is a special case of generalized chaotic synchronization, as in low-dimensional systems. Note that complete chaotic synchronization of systems described by coupled Ginzburg–Landau equations with different parameters is observed when the coupling is relatively strong.

Analogous results are obtained by calculating the largest spatial Lyapunov exponent $\Lambda_c$. Figure 3 illustrates the dependence of $\Lambda_c$ on the coupling strength $\varepsilon$ calculated by using the algorithm presented in Section 3 with $M = 1000$ and $T = 4$ in (7). One can clearly see that $\Lambda_c$ decreases with increasing $\varepsilon$ and vanishes when $\varepsilon = \varepsilon_{GS}$ (indicated by arrow in Fig. 3), which means the onset of a generalized chaotic synchronization regime. These results support the analysis performed above by using an auxiliary system: equal critical values $\varepsilon_{GS}$ of the coupling strength are obtained by both methods.

Now, let us discuss the mechanism leading to the onset of generalized chaotic synchronization in coupled spatially distributed systems.

In [63], we showed that the onset of generalized synchronization of low-dimensional chaotic dynamical systems depends on the behavior of the system modified by introducing additional dissipation into the response oscillatory system.

To analyze the onset of generalized chaotic synchronization in the case of spatially uniform dissipative coupling, we consider natural oscillations $v_m$ of the modi-

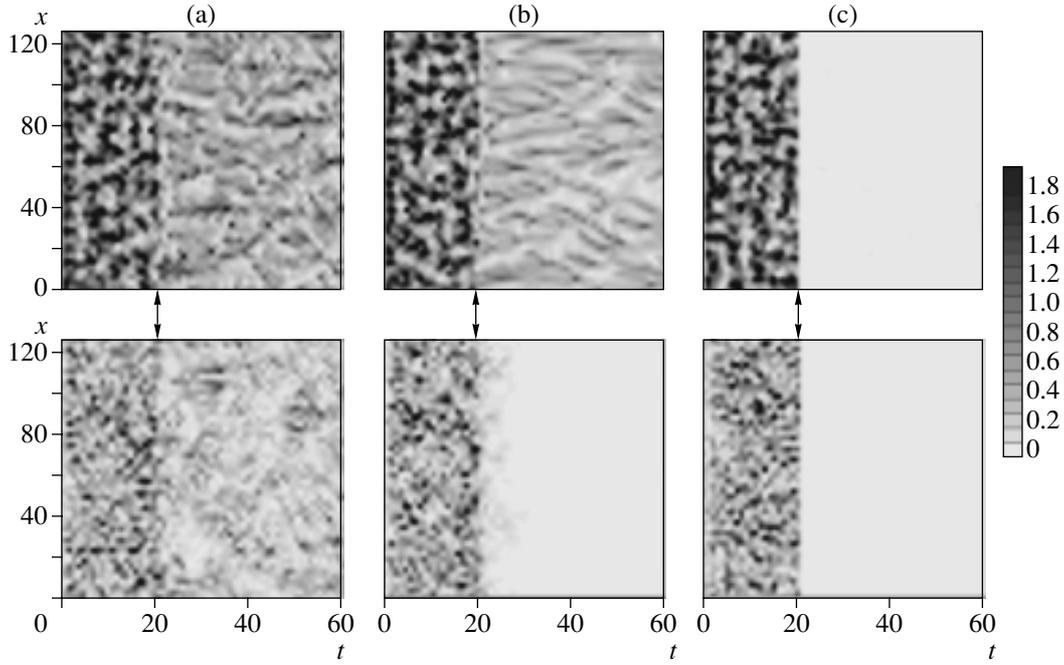

**Fig. 2.** Spatiotemporal evolution of amplitude difference between the driving and response systems, $|u(x, t) - v(x, t)|$ (top), and between the response and auxiliary systems, $|v(x, t) - v_a(x, t)|$ (bottom): (a) $\varepsilon = 0.5$, no generalized synchronization; (b) $\varepsilon = 0.9$, generalized synchronization; (c) $\varepsilon = 20$, complete generalized synchronization. Arrows indicate the instant when coupling is switched on ($t = \tau = 20$).

fied system derived from Eq. (2) by setting $u(x, t) = 0$ in (9):

$$\frac{\partial v_m}{\partial t} = v - (1 - i\alpha_r)|v_m|^2 v_m + (1 + i\beta_r)\frac{\partial^2 v_m}{\partial x^2} - \varepsilon v_m, \quad (10)$$
$$x \in [0, L].$$

It should be noted here that the term $-\varepsilon v_m(x, t)$ in (10) plays the role of additional dissipation.

The onset of generalized chaotic synchronization of dynamical systems is observed either when the value of $\varepsilon$ corresponds to transition between chaotic and regular (steady-state or periodic) regimes of the modified system or when the external forcing term $\varepsilon u_{ext}$ exceeds in amplitude natural oscillations $v_m$ of the modified system to the extent that the system's state is moved out of the attractor basin into a phase-space domain characterized by faster contraction of the phase-space volume (stronger dissipation). In the latter case, the natural dynamics of the system are suppressed, and the behavior of the response system is controlled by external forcing, as manifested by a functional relationship between the states of the driving and response systems [25, 63, 71].

In contrast to low-dimensional systems [63], transition of the coupled chaotic systems analyzed here to a generalized synchronization regime is simultaneously controlled by both mechanisms. A larger $\varepsilon$ corresponds to a larger amplitude of the external force on the one hand and stronger dissipation on the other (see also [63]). The latter effect reduces the amplitude of spatiotemporal oscillations in modified system (10), and a spatiotemporally uniform state ($|v_m(x, t)| = 0$) is reached at $\varepsilon = \varepsilon_0 = 1$. The spatiotemporally averaged squared oscillation amplitude $\langle v_m^2 \rangle$ plotted in Fig. 4a demonstrates that the oscillation intensity in the system described by the modified Ginzburg–Landau equation linearly decreases with increasing $\varepsilon$ because of the increase in the dissipative term $-\varepsilon v_m$.

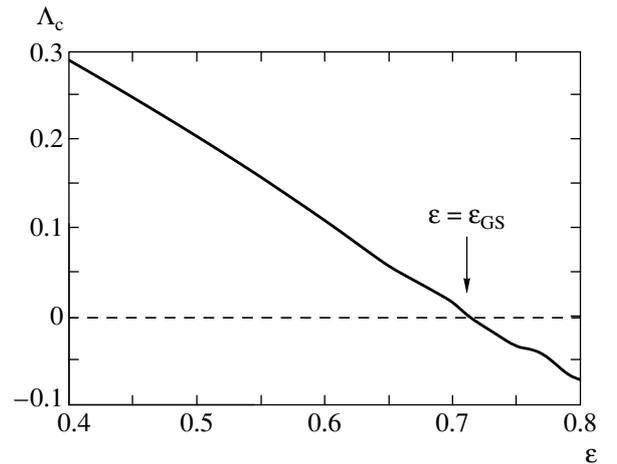

**Fig. 3.** Largest spatial conditional Lyapunov exponent for coupled system (1), (2) vs. coupling strength $\varepsilon$. Arrow indicates the onset of generalized synchronization at $\varepsilon = \varepsilon_{GS} \approx 0.73$.

However, the onset of a generalized synchronization regime is observed when $\varepsilon < \varepsilon_0 = 1$, owing to the second mechanism of generalized synchronization. Indeed, the spatiotemporally averaged squared amplitude $\langle(\varepsilon u)^2\rangle$ of the external forcing that drives the modified system plotted in Fig. 4a increases with $\varepsilon$, exceeding the averaged squared amplitude of natural oscillations of the modified system by a factor of about three at $\varepsilon = \varepsilon_{GS}$. In this case, the spatiotemporal state of the modified system is moved into a phase-space domain of strong dissipation, where the natural spatiotemporal chaotic dynamics of the modified system are suppressed. This leads to the onset of a generalized synchronization regime at $\varepsilon_{GS} < \varepsilon_0$. It is important that both mechanisms described above contribute to the onset of generalized synchronization when the coupling strength lies in the interval $(\varepsilon_{GS}, \varepsilon_0)$.

The last remark means that the onset of a generalized synchronization regime depends on parameters of modified system (10), but not on those of the driving system. Figure 4b demonstrates that $\varepsilon_{GS}$ varies only with the parameters $\alpha_r$ and $\beta_r$ of the response system in a wide range of control parameters, being virtually independent of the parameter $\beta_d$ of driving system (1). These results validate the analysis of generalized synchronization using the modified system [49, 63].

## 5. PINNING COUPLING

Now, we consider the case when the driving and response systems are coupled only at a finite number of points separated by a distance $\Delta X$:

$$\mathcal{F}[u, v] = \delta(x - N\Delta X)(u - v), \quad N = 0, 1, 2, \ldots, \quad (11)$$

where $\delta(\xi)$ is the Dirac delta function and the $\Delta X$ can be regarded as the length scale of spatially nonuniform coupling (the case of uniform coupling examined in Section 4 corresponds to $\Delta X = 0$).

Pinning coupling at a finite number of points in the domain of distributed systems was previously examined in analyses of complete chaotic synchronization of identical distributed systems [51, 56] and in studies of control of systems exhibiting chaotic behavior [57, 43].

We examine this type of coupling for the values of control parameters of the response and driving systems equal to those used in the preceding section. Figure 5 illustrates the evolution of the difference $|v(x, t) - u(x, t)|$ between the states of the response and driving systems and the difference $|v(x, t) - v_a(x, t)|$ between the states of the response and auxiliary systems for $\Delta X = 7.36$ and several values of $\varepsilon$. When the coupling

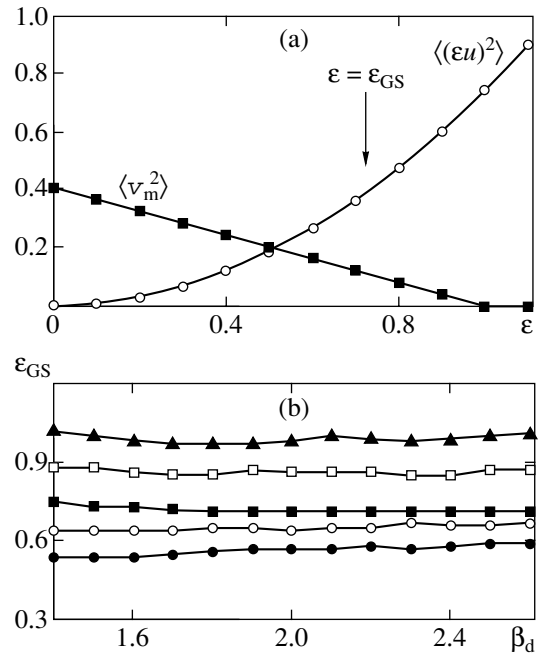

**Fig. 4.** (a) Averaged squared amplitudes of modified-system oscillation and external forcing that drives the response system vs. coupling strength. (b) Threshold coupling strength $\varepsilon_{GS}$ corresponding to the onset of a generalized synchronization regime vs. control parameter $\beta_d$ of the driving system: $\alpha_r = \beta_r = 3.0$ (●), 3.5 (○), 4.0 (■), 5.0 (□), 6.0 (▲).

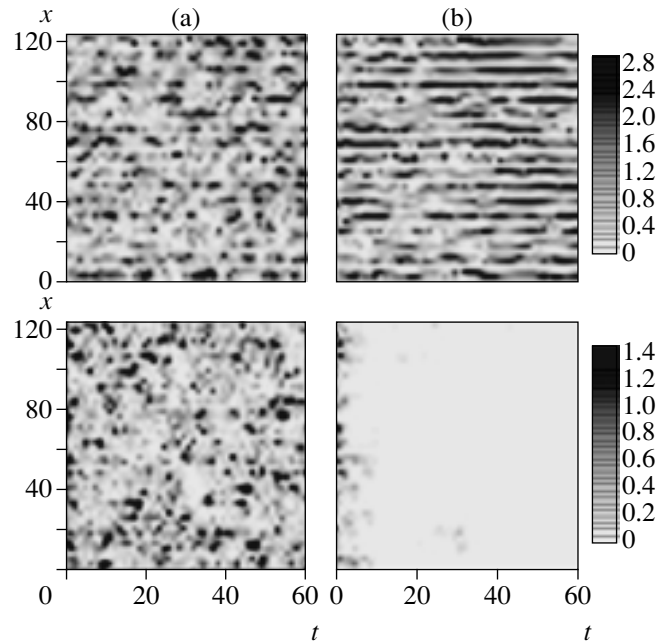

**Fig. 5.** Spatiotemporal evolution of amplitude difference between the driving and response systems, $|u(x, t) - v(x, t)|$ (top), and between the response and auxiliary systems, $|v(x, t) - v_a(x, t)|$ (bottom) for pinning coupling with $\Delta X = 7.36$: (a) $\varepsilon = 20$, no synchronization; (b) $\varepsilon = 100$, generalized synchronization of distributed oscillatory systems.

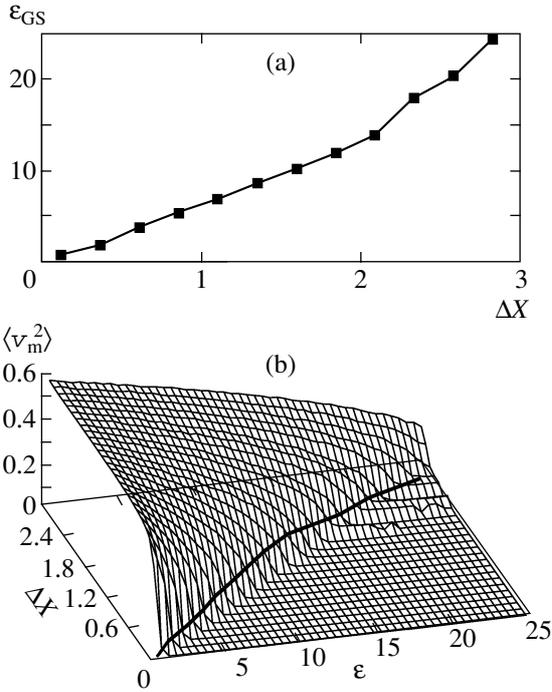

**Fig. 6.** (a) Generalized synchronization threshold $\varepsilon_{GS}$ vs. distance $\Delta X$ between coupling points. (b) Spatiotemporally averaged amplitude of modified-system oscillations vs. $\varepsilon$ and $\Delta X$; bold curve corresponds to generalized synchronization threshold.

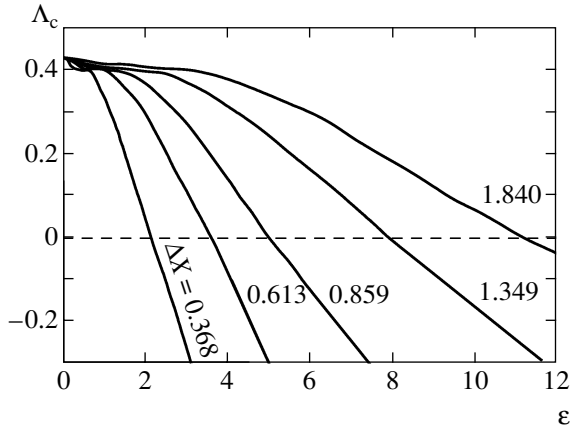

**Fig. 7.** Largest spatial conditional Lyapunov exponent for system (1), (2) with pinning coupling vs. coupling strength for several values of $\Delta X$.

strength is relatively small ($\varepsilon = 20$), the response and auxiliary systems are in different states (see Fig. 5a); i.e., no generalized synchronization is observed. The onset of generalized synchronization in spatially distributed systems with stronger localized coupling is illustrated by Fig. 5b ($\varepsilon = 100$), where identical spatiotemporal states of the response and auxiliary systems ($v(x, t) \equiv v_a(x, t)$) observed after the initial transients have died out correspond to generalized synchronization of systems with pinning coupling.

It was shown in the previous section that the generalized chaotic synchronization threshold $\varepsilon_{GS}$ weakly depends on the control parameters of the driving system in the case of spatially uniform coupling. Analogous results were obtained in the case of a pinning coupling, except that the threshold $\varepsilon_{GS}$ depends on the degree of nonuniformity of the coupling (the distance $\Delta X$ between the coupling points), as illustrated by Fig. 6a, where $\varepsilon_{GS}$ increases with $\Delta X$ from $\varepsilon_{GS} \approx 0.76$ at $\Delta X = 0$.

Let us analyze the onset of generalized chaotic synchronization in the case of localized coupling by the modified-system method. In the case of spatially localized diffusion coupling, the modified system is written as

$$\frac{\partial v_m}{\partial t} = v_m - (1 - i\alpha_r)|v_m|^2 v_m + (1 + i\beta_r)\frac{\partial^2 v_m}{\partial x^2} + \varepsilon \mathcal{F}[0, v_m], \quad (12)$$
$$x \in [0, L],$$

where $\mathcal{F}$ is defined by (11).

Let us examine the dynamics of modified system (12) by varying the strength $\varepsilon$ of the nonuniform coupling. Figure 6b shows the spatiotemporally averaged oscillation intensity $\langle v_m^2 \rangle$ plotted as a function of [epsilon] in $\Delta X$ for the modified system. As in the case of spatially uniform coupling (see also [49]), the onset of generalized synchronization is observed when the amplitude of chaotic oscillations decreases with increasing coupling strength $\varepsilon$ for each particular value of $\Delta X$ (see Fig. 6b). With increasing $\varepsilon$, the value of $\varepsilon^2$ substantially exceeds $\langle v_m^2 \rangle$. An increase in $\Delta X$ (i.e., a decrease in the number of coupling points) leads to a decrease in the rate of additional dissipation. As a consequence, the onset of generalized synchronization is observed at a higher $\varepsilon_{GS}$ (see Fig. 6a and bold curve in Fig. 6b). However, the average oscillation intensity $\langle v_m^2 \rangle$ at the threshold $\varepsilon_{GS}$ corresponding to different $\Delta X$ is constant in a wide range of parameters.

Figure 7 shows the largest spatial conditional Lyapunov exponent plotted versus $\varepsilon$ for several values of $\Delta X$. It is clear that the exponent calculated for a particular value of $\varepsilon$ increases with $\Delta X$; i.e., the response system becomes increasingly unstable. Since the onset of generalized synchronization corresponds to zero value of $\Lambda_c$, the value of $\varepsilon_{GS}$ must increase with $\Delta X$, as illustrated by Fig. 6a.

Thus, generalized chaotic synchronization is also observed in distributed systems with pinning coupling. As in the case of diffusion coupling, the corresponding generalized synchronization threshold is determined by the dynamics of modified Ginzburg–Landau equa-tion (12), where additional dissipation is localized at a finite number of points in space.

## 6. DIFFUSION COUPLING IN A BOUNDED REGION: PARTIAL GENERALIZED SYNCHRONIZATION

Now, we consider the case of diffusion coupling localized within a region $\tilde{S}$ (an interval of width $D$). In the rest of the domain (an interval of width $L - D$), the distributed systems remain uncoupled. The corresponding coupling function is

$$\mathcal{F}[u, v] = H\left(x - \frac{L-D}{2}\right) \times H\left(\frac{L+D}{2} - x\right)(u - v), \quad (13)$$

where $H$ is the Heaviside step function.

An analysis shows that the onset of a functional relationship between the states of the driving and response systems is observed only within a subregion $S_{sync} \in S$ as $\varepsilon$ is increased.

Thus, we can define *partial generalized synchronization* of spatially nonuniform distributed oscillatory systems as a new type of chaotic synchronization characterized by the onset of a functional relationship between oscillations within bounded parts $S_{sync}$ of the systems' domains.

Partial generalized chaotic synchronization can be detected by the auxiliary system method as the onset of identical oscillations within a bounded region in the domain of the response and auxiliary systems.

In the case considered here, a regime of this type is illustrated by Fig. 8, where the difference in oscillation amplitude between the response and auxiliary systems is plotted for several values of coupling strength. Here, partial generalized synchronization manifests itself by a functional relationship between $v_a(x, t)$ and $v(x, t)$ in a region $S_{sync}$, while the oscillations of the response and auxiliary systems are not synchronized in the rest of the domain (in regions $\tilde{S}$ and $S_{async} = S - S_{sync}$). It is obvious that boundary effects play an essential role at the periphery of $S$ in the case of a highly nonuniform coupling described by (13).

When the value of $\varepsilon$ is increased while the width of the coupling region is held constant (see Fig. 8), synchronization is not observed so long as the coupling strength remains relatively small. With further increase in $\varepsilon$, the onset of partial generalized synchronization is observed (see Fig. 8b) within a subregion $S_{sync}$ of $S$ (its central part, where boundary effects are minimal). The width of the $S_{sync}$ increases with $\varepsilon$ (see Figs. 8c and 8d).

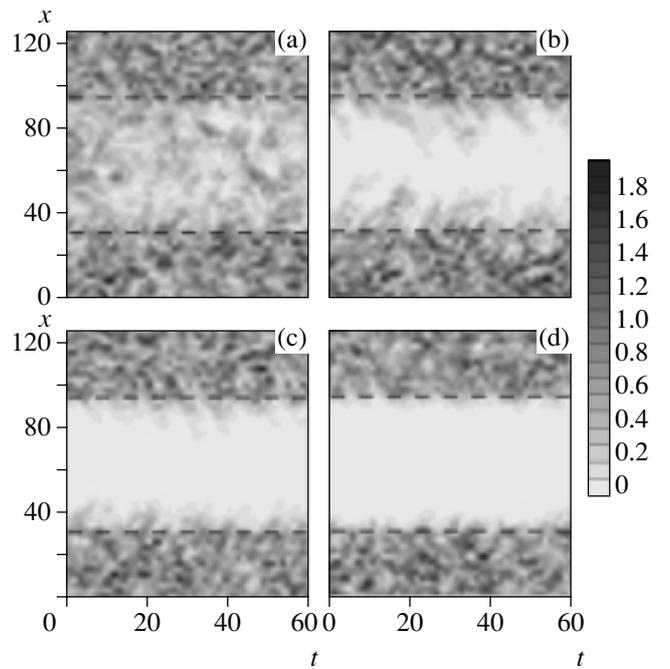

**Fig. 8.** Spatiotemporal evolution of the amplitude difference between the response and auxiliary systems, $|v(x, t) - v_a(x, t)|$, in the case of dissipative coupling within $D = 0.5L$ (between dashed lines). Transition from unsynchronized to partial generalized synchronization regime: (a) $\varepsilon = 0.6$; (b) $\varepsilon = 0.825$; (c) $\varepsilon = 1.0$; (d) $\varepsilon = 1.6$.

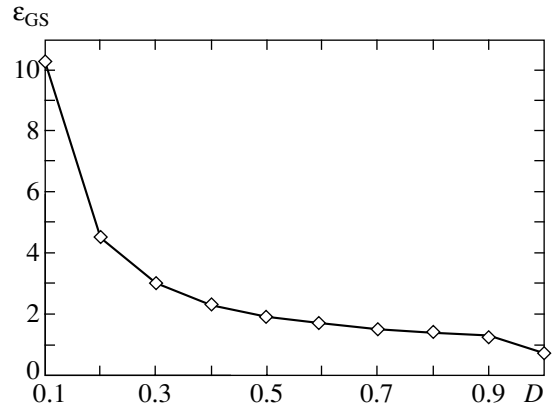

**Fig. 9.** Threshold coupling strength $\varepsilon_{GS}$ corresponding to the onset of partial generalized synchronization vs. width $D$ of the coupling region.

However, the difference between oscillations of the response and auxiliary systems due to boundary effects persists at the periphery of $S$; i.e., $S_{sync}$ will remain smaller than $S$.

Figure 9 shows the threshold coupling strength $\varepsilon_{GS}$ for partial generalized synchronization plotted versus width $D$ of the coupling region $S$. The rapid decrease in $\varepsilon_{GS}$ with increasing $D$ is explained by the stronger influence of boundary effects on the dynamics driven by chaotic external forcing in the case of a narrower coupling region.

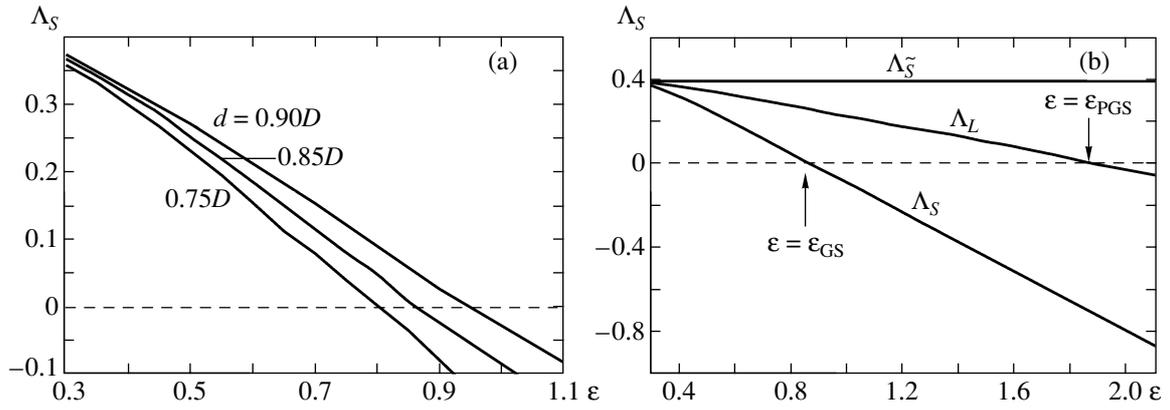

**Fig. 10.** Largest spatial conditional Lyapunov exponents vs. coupling strength: (a) $\Lambda_S$ calculated for several subregions of width $d$ in the coupling region; (b) $\Lambda_{\tilde{S}}$, $\Lambda_S$, and $\Lambda_L$ for different regions. Arrows indicate threshold values.

Partial generalized synchronization can also be detected by calculating spatial conditional Lyapunov exponents for some regions. Since there are two regions of essentially different dynamics ($S$ and $\tilde{S}$) in the case considered here, the respective Lyapunov exponents can be evaluated separately by using either region as the domain of integration in (6).

Let us consider the largest conditional Lyapunov exponent evaluated for the coupling region $S$. With increasing $\varepsilon$, the onset of a partial generalized synchronization is observed: a region $S_{\text{sync}} \in S$ of functional relation between the oscillations in the driving and response systems appears. The largest spatial conditional Lyapunov exponent $\Lambda_S$ calculated for $S_{\text{sync}}$ becomes negative with increasing coupling strength, which signifies the onset of partial generalized synchronization. At the same time, the value of $\Lambda_{\tilde{S}}$ calculated for the rest of the domain ($\tilde{S}$ and $S_{\text{async}}$) remains positive. The onset of a partial generalized synchronization regime with increasing coupling strength is illustrated by Fig. 10a, where $\Lambda_S(\varepsilon)$ is shown for several values of the width $d \leq D$ of the region where the exponent is evaluated.

Thus, when partial generalized synchronization in nonuniformly coupled systems is detected by calculating conditional Lyapunov exponents, the coupling strength corresponding to the onset of partial generalized synchronization is determined up to an uncertainty due to boundary effects at the periphery of $S$. In other words, the application of Lyapunov exponents to spatially nonuniform systems is a difficult task requiring a careful analysis.

We define the integral largest spatial conditional Lyapunov exponent for the entire domain of the system as a linear combination of the spatial Lyapunov exponents evaluated for the coupled and uncoupled regions:

$$\Lambda_L = \mu(S)\Lambda_S + \mu(\tilde{S})\Lambda_{\tilde{S}}, \quad (14)$$

where $\mu(x)$ is the measure (in the present case, width) of a particular region ($\mu(S) + \mu(\tilde{S}) = 1$).

It is obvious that $\Lambda_{\tilde{S}} = \text{const} > 0$ for any $\varepsilon$. As shown above, $\Lambda_S$ decreases with increasing $\varepsilon$. If

$$|\Lambda_S(\varepsilon)| > \mu(\tilde{S})\Lambda_{\tilde{S}}/\mu(S), \quad \Lambda_S(\varepsilon) < 0,$$

then $\Lambda_L < 0$, as illustrates by Fig. 10b, where $\Lambda_S$, $\Lambda_{\tilde{S}}$, and $\Lambda_L$ defined by (14) are plotted as functions of $\varepsilon$. The zero value of $\Lambda_S$ at $\varepsilon = \varepsilon_{GS}$ corresponds to the onset of generalized chaotic synchronization in the coupling region. Note that $\Lambda_L(\varepsilon_{GS}) > 0$. However, there exists a larger value of coupling strength, $\varepsilon = \varepsilon_{PGS}$, such that $\Lambda_L(\varepsilon_{PGS}) = 0$. This means that the behavior of the response system as a whole is determined by the dynamics of the driving system, but one-to-one correspondence is observed only in a bounded region $S$, as demonstrated by the auxiliary system method.

Thus, partial generalized chaotic synchronization observed in distributed oscillatory systems with uniform diffusion coupling localized in a bounded region is characterized by a functional relationship between the oscillations localized within a bounded subregion (where the oscillations of the response and auxiliary systems are identical). In the rest of the domain, there is no functional relationship between the oscillations; i.e., the response and auxiliary systems exhibit different spatiotemporal dynamics. Analysis of generalized synchronization based on the use of the largest conditional Lyapunov exponent is difficult to apply because of the boundary effects at the periphery of the coupling region.

# 7. GENERALIZED SYNCHRONIZATION IN GINZBURG–LANDAU EQUATIONS IN THE PRESENCE OF FLUCTUATIONS

The operation of nonlinear oscillatory systems in radio engineering, physiology, chemistry, etc. is affected by various fluctuations. Therefore, of special interest is robustness to noise in generalized chaotic synchronization regimes of distributed oscillatory systems.

To examine the effects due to fluctuations, we consider unidirectionally coupled Ginzburg–Landau equations perturbed by spatially distributed white noise:

$$\frac{\partial u}{\partial t} = u - (1 - i\alpha_d)|u|^2 u + (1 + i\beta_d)\frac{\partial^2 u}{\partial x^2} + \tilde{D}\zeta(x, t), \quad (15)$$
$$x \in [0, L],$$

$$\frac{\partial v}{\partial t} = v - (1 - i\alpha_r)|v|^2 v + (1 + i\beta_r)\frac{\partial^2 v}{\partial x^2} + \tilde{D}\zeta(x, t) + \varepsilon H(t - \tau)\mathscr{F}[u, v], \quad (16)$$
$$x \in [0, L],$$

where $\zeta(x, t)$ is a delta-correlated zero-mean Gaussian process,

$$\langle \zeta(x, t) \rangle = 0,$$
$$\langle \zeta(x, t)\zeta(x', t') \rangle = \delta(x - x')\delta(t - t'), \quad (17)$$

and $\tilde{D}$ is the noise intensity. The simplest case of spatially uniform diffusion coupling is considered: $F[u, v] = u - v$.

A numerical analysis of the Ginzburg–Landau equation with an additional stochastic term was performed by using the following standard scheme for integrating stochastic partial differential equations [72]. The complex variable $u(x, t)$ is represented as a complex field defined on a one-dimensional grid with mesh size $\Delta x$: $u_i(t) = u(x_i, t)$ is defined at $x_i = i\Delta x$ ($i = 1, \ldots, N$). Then, a delta-correlated zero-mean Gaussian process $\zeta_i(t)$ is defined by the condition

$$\langle \zeta_i(t)\zeta_j(t') \rangle = \tilde{D}\delta_{ij}\delta(t - t')/\Delta x,$$

where $D = \tilde{D}/\Delta x$ is the intensity of spatiotemporal noise in a discrete space [72]. The Laplacian operator is approximated by a three-point finite-difference scheme [73]. Thus, stochastic partial differential equations (15) and (16) are computed as one-dimensional lattice of maps by the one-step Euler method with time step $\Delta t$.

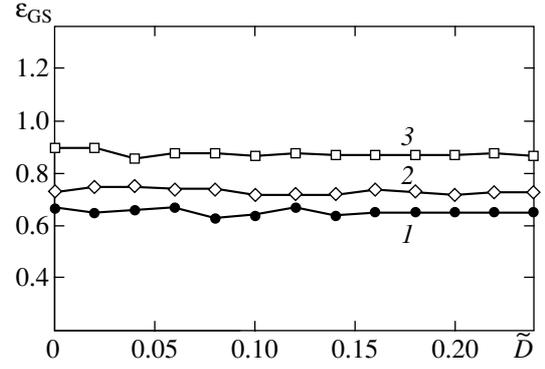

**Fig. 11.** Threshold coupling strength corresponding to the onset of generalized synchronization vs. noise level $\tilde{D}$: (*1*) $\alpha_r = 3$, $\beta_r = 3$; (*2*) $\alpha_r = 4$, $\beta_r = 4$; (*3*) $\alpha_r = 5$, $\beta_r = 5$.

Generalized synchronization was detected by the auxiliary system method with an auxiliary system perturbed by noise similar to that in the response system. In the presence of fluctuations, we use the following criterion for detecting generalized synchronization.

Generalized synchronization is supposed to take place when the mean square difference between the states of the driving and auxiliary systems satisfies the condition

$$\frac{1}{T}\int_T\int_0^L |v(x, t) - v_a(x, t)|^2 < \delta, \quad (18)$$

where $\delta$ is set equal to 0.01.

Figure 11 shows the threshold $\varepsilon_{GS}$ for the onset of generalized synchronization plotted versus the noise intensity $\tilde{D}$ for several values of control parameters. It is clear that low-intensity noise almost does not affect the threshold for generalized chaotic synchronization in distributed systems. As the noise intensity $\tilde{D}$ exceeds 0.5, the threshold value of coupling strength begins to increase monotonically. This behavior of a noisy system in generalized chaotic synchronization regimes is explained by the fact that the generalized synchronization threshold is primarily determined by the properties of the modified system, as shown above. Low-intensity zero-mean noise does not have any significant effect on the behavior of the modified system and, therefore, on the generalized chaotic synchronization threshold. However, high-intensity noise can significantly change the dynamical characteristics and even the dynamical regime of the modified system. Accordingly, the threshold coupling strength $\varepsilon_{GS}$ changes in the presence of noise. The fact that low-intensity noise almost does not change the generalized synchronization threshold suggests that the phenomenon under analysis is robust in coupled distributed oscillatory systems.

## 8. CONCLUSIONS

We analyze generalized synchronization in distributed oscillatory systems exhibiting spatiotemporal chaotic behavior described by Ginzburg–Landau equations. Generalized chaotic synchronization in a spatially distributed system is detected both by the auxiliary system method used in [49] to analyze generalized synchronization of a distributed system and the largest spatial conditional Lyapunov exponent introduced in this study.

Generalized synchronization is examined for several types of coupling between the systems. For all types of coupling, the onset of generalized synchronization is due to suppression of natural spatiotemporal chaotic oscillations by additional dissipation in the medium. In particular, the natural oscillation intensity is reduced by introducing an additional dissipative term into the complex Ginzburg–Landau equation. Furthermore, natural dynamics of the response system are suppressed by the increase in external forcing amplitude with coupling strength, because the state of the system is moved into a phase-space domain of strong dissipation. These effects are examined by analyzing a modified Ginzburg–Landau equation with additional dissipation.

An analysis of the behavior of a spatially nonuniform system with coupling localized within a bounded region of the system's domain has revealed a new type of chaotic synchronization called partial generalized synchronization. It should be noted that boundary effects strongly affect synchronized dynamic behavior of systems at the periphery of the coupling region.

## ACKNOWLEDGMENTS

We thank D.I. Trubetskov for interest in this study and helpful critical remarks. This work was supported by the Russian Foundation for Basic Research (project nos. 05-02-16273 and 06-02-16451), under the Federal Scientific and Technical Program "Research and Development in Priority Areas of Science and Technology" (grant nos. NIR 2006-RI-19.0/001/053 and 2006-RI-19.0/001/054), and under the Program of Support for Leading Science Schools of the Russian Federation (grant no. 4167.2006.2). We also gratefully acknowledge the support provided by the Dynasty Foundation affiliated with the International Center for Fundamental Physics in Moscow.